\documentclass[onecolumn,showpacs,preprintnumbers,amsmath,amssymb,11pt]{revtex4}

\usepackage{epsf}
\usepackage{graphicx}  
\usepackage{dcolumn}   
\usepackage{bm}        

\newcommand{\be}{\begin{equation}}
\newcommand{\en}{\end{equation}}
\newcommand{\bea}{\begin{eqnarray}}
\newcommand{\ena}{\end{eqnarray}}
\begin{document}

\preprint{}

\title{Thermodynamical aspects of modified holographic dark energy model}

 \author{Hui Li \footnote{Electronic address: lihui@ytu.edu.cn} }
 \affiliation{
 Department of Physics, Yantai University, 30 Qingquan Road, Yantai 264005, Shandong Province, P.R.China }

\author{Yi Zhang\footnote{Electronic address: zhangyia@cqupt.edu.cn}}
\affiliation{College of Mathematics and Physics, Chongqing University of
Posts and Telecommunications, Chongqing 400065, P.R.China}

\begin{abstract}

We investigate the unified first law and the generalized second law in a modified holographic dark energy model. The thermodynamical analysis on the apparent horizon can work and the corresponding entropy formula is extracted from the systematic algorithm. The entropy correction term depends on the extra-dimension number of the brane as expected, but the interplay between the correction term and the extra dimensions is more complicated. With the unified first law of thermodynamics well-founded, the generalized second law of thermodynamics is discussed and it's found that the second law can be violated in certain circumstances. Particularly, if the number of the extra dimensions is larger than one, the generalized law of thermodynamics is always satisfied; otherwise, the validity of the second law can only be guaranteed with the Hubble radius greatly smaller than the crossover scale $r_c$ of the $5$-dimensional DGP model.

\end{abstract}


\maketitle

\section{Introduction}

Our universe endures an accelerating expansion. A hypothetic dark energy component in our universe is suspected to take charge of the unexpected kinematics\cite{synthesis}. Following this approach, different strategies have been developed to understand the details about its value and the  coincidence with that of the other dark matter component. Confronting with the unsatisfactory argument of the dark energy field theoretical peculiarity, people tried to test the modification of gravity theories and certain speculative quantum-gravity principles, e.g. the holographic principle.

The holographic principle in quantum gravity was originally proposed by G. 't Hooft inspired mostly by blackhole physics and has obtained great enthusiasm in theoretical physics with the promotion of its recent successful realization--AdS/CFT correspondence\cite{holography}. Due to the holographic property of gravity, local quantum field theory could be invalided at energy scale far less than the traditional effective field theory predicts. By assuming the energy in a box to be less than the mass of a black hole with the same size, a stringent constraint between the IR cutoff length scale and the UV cutoff energy scale reproduces the so-called  UV/IR relation in quantum gravity\cite{Cohen}. Rough estimation shows that, if we choose the IR cutoff length scale to be of the order of the Hubble radius in cosmology, the corresponding UV cutoff energy scale meets the magnitude of present observed dark energy well, which could hardly be an accidental coincidence without fundamental reasons. Following this approach, many instructive proposals on the dark energy construction have emerged in the past decade; for instance, the holographic dark energy model and its invariants\cite{HDE}, agegraphic dark energy models\cite{age} and holographic Ricci dark energy\cite{Ricci}, etc..
In this paper, we will focus on the modified holographic  dark energy model \cite{Gongli} \cite{Gongli2} where  the mass of the black hole is the cutoff.  Besides the utility of the higher-dimensional blackhole relation of mass to radius, this theoretical construction contains the DGP model when $N=4 $\cite{DGP}, and the  LCDM concordance model when $N=5$.  The thermodynamics in DGP and LCDM is clearly different, and it is worth further refinement of the thermodynamical behavior with the varied $N$ which may disclose some implicit relation of spacetime dimensions\cite{Vil} to the formulation of the thermodynamical laws.

Below we will firstly give a brief introduction to the higher-dimensional dark energy model with Hubble radius the IR cutoff. Then the thermodynamical analysis based on the systematic algorithm elaborated in the work of Cai, et.al. will be carried out\cite{Cai}. After assuming the foundation of the unified first law of thermodynamics\cite{Hayward}, the corrected entropy formula\cite{Shek1} deviating from the simple area-entropy one will be obtained. Therefore the generalized second law of thermodynamics can be formulated and investigated in details\cite{Shek2}. The discussions and conclusion will appear in the end.

\section{The Modified Holographic Model }

 Various competitive candidates of dark energy were discussed in the holographic viewpoint, including the holographic dark energy with the event horizon the IR cutoff, the holographic Ricci dark energy with the Ricci scalar the IR cutoff, and the age-graphic dark energy with the age of the universe the cutoff, etc.. A more recent model of this kind was the so-called modified holographic dark energy with the mass of the Schwarzschild black hole as cutoff and the UV/IR mixing stemming from the higher dimensional version of the original Cohen's proposal. In $N+1$ bulk dimensions with $N$ the spatial dimension and $n\equiv N-3$ the extra dimension, the dark energy density and the pressure in the modified holographic dark energy model are\cite{Gongli}
\begin{eqnarray}
\label{mholrhod}
&&\rho_d= \frac{d(N-1)\Omega_{N-1}}{16\pi G_N}H^{5-N},\\
\label{yaqiang}
&&p_{d}=-\rho_{d}+\frac{d(N-5)(N-1)\Omega_{N-1}}{48\pi G_N}H^{3-N}\dot{H}.
\end{eqnarray}
where the dimensionless parameter $d$ is naturally of order one and encodes some theoretical adjustment. Following the work of Ref.\cite{Gongli} \cite{Gongli2}, throughout the paper the specified IR cutoff is simply the Hubble radius of the observed universe. Then, the corresponding first Friedmann equation is
\begin{equation}
\label{fried1}
H^2=\frac{1}{3M_{pl}^2}[\rho_m+\frac{d(N-1)\Omega_{N-1}}{16\pi G_N}H^{5-N}].
\end{equation}
Here, we assume the energy conservation law of the individual dark energy component and the matter, which gives
\begin{eqnarray}
&&\dot{\rho}_{d}+3 H(\rho_{d}+p_{d})=0,\\
\label{continuationd}
&&\dot{\rho}_m+3 H(\rho_m+p_m)=0.
\label{continuation}
\end{eqnarray}
where the subscript $m$ denotes the quantities of matter in the universe throughout the paper by default. Therefore, with the effective evolution of the modified holographic dark energy included, the second Friedmann equation can be easily obtained:
\begin{equation}
\label{fried2}
2\dot{H}+3H^2=-\frac{1}{M_{pl}^2}[p_m+\frac{(N-5)(N-1)\Omega_{N-1}d}{48\pi G_N}H^{3-N}\dot{H}-\frac{(N-1)\Omega_{N-1}d}{16\pi G_N}H^{5-N}].
\end{equation}

\section{the First Law of Thermodynamics}
As was implicitly assumed in the above two Friedmann equations, the metric used $g_{\mu\nu}$ is
\begin{eqnarray}
\nonumber
ds^2=-dt^2+\frac{a(t)^2}{1-kr^2}dr^2+a(t)^2r^2d\Omega_{n-2}^2=h_{ab}dx^a dx^b+\tilde{r}^2d\Omega_{n-2}^2,
\end{eqnarray}
where $\tilde{r}=a(t)r$, $x^0=t, x^1=r$, $h_{ab}=diag(-1,a^2/(1-kr^2))$ and $k$ denotes the spatial curvature.

For a spherically symmetric  spacetime, the Einstein equation can be reformulated to be the form of the unified first law of thermodynamics:
\begin{equation}
\label{unified}
dE=A\Psi+WdV
\end{equation}
with $E$ the Misner-Sharp energy\cite{Misner}, $A$ the area of the sphere, $V$ the volume, $\Psi$ the energy supply vector and $W$ the work density.
This reformulation can certainly apply for the FRW cosmological context.
Following Ref.\cite{Hayward}, the energy -supply vector $\Psi$ and the work density can be defined as
\begin{equation}
\Psi_a=T^b_a\partial_b \tilde{r}+W\partial_a\tilde{r},W=-\frac{1}{2}T^{a b}h_{a b}.
\end{equation}
where $T^{a b}$ is the projection of the $(3+1)$-dimensional energy-momentum tensor $T^{\mu\nu}$ of a cosmological energy component in the normal direction of $2$-sphere. For the present case, it is easy to find
\begin{equation}
\label{vectors}
\Psi=-\frac{1}{2}(\rho+p)H \tilde{r} dt+ \frac{1}{2}(\rho+p)a dr,W=\frac{1}{2}(\rho-p)
\end{equation}

In FRW cosmology, modified gravity theories can always be regarded to be composed of two parts with one of them the ordinary matter contribution and the other the effective component. Therefore, the energy-supply vector can be written to be a splitting form\cite{Cai}:
\begin{equation}
\label{split}
\Psi=\Psi_m+\Psi_d.
\end{equation}
Suppose that the energy-momentum tensor $T^{\mu\nu}$ of matter has the same form of a perfect fluid
$T^{\mu\nu}=(\rho+p)U^{\mu}U^{\nu}+p g^{\mu\nu}$ as that of the effective dark energy with $\rho$ and $p$ are the corresponding energy density and pressure respectively.
Then Equations (\ref{vectors}) and (\ref{split}) indicate that
\begin{eqnarray}
&&\Psi_m=-\frac{1}{2}(\rho+p)H\tilde{r}dt+\frac{1}{2}(\rho+p) a dr,\\
\label{psie}
&&\Psi_d=-\frac{1}{2}(\rho_{d}+p_{d})H\tilde{r}dt+\frac{1}{2}(\rho_{d}+p_{d}) a dr.
\end{eqnarray}

To get the ordinary first thermodynamics law and to calculate traditional thermodynamical quantities, there is strong evidence in establishing the true first law of thermodynamics on the apparent horizon with Equation (\ref{unified}). The dynamical apparent horizon is determined by the equality $h^{ab}\partial_a \tilde{r}\partial_b \tilde{r}=0$, and its radius can be found out:
\begin{eqnarray}
\tilde{r}_A=\frac{1}{H}\label{rA}, \,\,\,\,\dot{\tilde{r}}_A=-\tilde{r}^2_A\dot{H}.\label{rA2}
\end{eqnarray}
for the spatially flat $k=0$ case. That is to say, the apparent horizon coincides with the Hubble horizon in spatially flat FRW case.
As a matter of fact, after projecting it along a vector $\xi=\partial_t-(1-2\epsilon)Hr\partial_r$ tangential to the apparent horizon with
$\epsilon=\dot{\tilde{r}}_A/(2H\tilde{r}_A$), the true first law of thermodynamics of the apparent horizon is obtained~\cite{Cai}:
\begin{equation}
\langle dE, \xi\rangle=\frac{\kappa}{8\pi G}\langle dA,
\xi\rangle+\langle WdV, \xi\rangle\,
\end{equation}
with $\kappa=-(1-\dot {\tilde r}_A/(2H\tilde r_A))/\tilde r_A$ the surface gravity of the apparent horizon.
The pure matter energy-supply $A\Psi_m$ (after projecting on the apparent horizon) gives the heat flow $\delta Q$ in the Clausius
relation $\delta Q=T$d$S$. By using the unified first law on the apparent horizon, we have
\begin{equation}
\delta Q\equiv \langle A\Psi_m,\xi \rangle= \frac{\kappa}{8\pi G}\langle dA,\xi \rangle-\langle A\Psi_d,\xi\rangle
\label{Apsim}
\end{equation}
From Equations ~(\ref{mholrhod}), ~(\ref{yaqiang}) and ~(\ref{psie}), the above equality can be recast into
\begin{equation}
\langle A\Psi_m,\xi \rangle=T\langle \frac{2\pi\tilde{r}_A d\tilde{r}_A}{G}+\frac{\pi d(N-5)(N-1)\Omega_{N-1}\tilde{r}_A^{N-2}}{6G_N}d\tilde{r}_A,\xi \rangle
\label{ds}
\end{equation}
with $8\pi G=1/M^2_{pl}$ and the temperature of the apparent horizon\cite{apparent}
\begin{equation}
T=\frac{\kappa}{2\pi}.
\label{T}
\end{equation}
Compared with the Clausius relation $\delta Q=T dS$, it is formally found that
\begin{equation}
\label{tds}
TdS=\frac{\kappa}{2\pi}[\frac{2\pi \tilde{r}_A}{G}d\tilde{r}_A+\frac{\pi d(N-5)(N-1)\Omega_{N-1}\tilde{r}_A^{N-2}}{6G_N}d\tilde{r}_A]
\end{equation}
and the above integration is well prepared for accomplishment. After integration, the entropy differential associated with the matter flowing through the apparent horizon in the time interval $dt$ and being inside the horizon determines the full entropy-area relation which is not merely the usual Bekenstein-Hawking entropy. This can be an entropy expression associated with the apparent horizon of an FRW universe described by the modified Friedmann equations by using the method proposed in Ref.~\cite{Cai}. Specifically, the corresponding entropy scaling which deviates from the usual $S=A/(4G)$ and is therefore a corrected entropy-area relation in the modified holographic dark energy model:
\begin{equation}
\label{s}
 S=\frac{A}{4G}+\frac{\pi d (N-5)\Omega_{N-1}}{6G_N}\tilde{r}_A^{N-1}
\end{equation}
with the proper apparent horizon area $A=4\pi \tilde{r}_A^2$.

 Not surprisingly, the entropy of the higher-dimensional dark energy model does not obey the usual area law on the brane and the correction term lies in the reduction of high dimensional bulk effect on the D-$3$ brane and depends quantitatively on the dimension of the bulk where the brane is embedded. It's worth noting that, instead of the factor $2$ in the second term of previous work(for instance, see \cite{Shek1}), the coefficient $(N-5)$ exhibits a new effect on the Bekenstein-Hawking entropy of the $4$-D Einstein gravity where the Einstein-Hilbert term on the brane contributes to this area term. Since there is no $Z_2$ symmetry property of the bulk to be  respected and the extra dimensions can be larger than one, the thermodynamical analysis emphasizes on a particular parameter characterizing the signature change of the entropy correction with the variation of the bulk dimension. When the setting-up of the present model refers to the co-dimension one brane-world, the entropy correction term becomes negative and tends to decrease the entropy corresponding to the brane gravity. Apart from that, the entropy-area relation takes the same form as that of various DGP brane-world models well studied in the literature. Even though the other branch of the DGP brane-world can be chosen and the first Friedmann equation takes the form different from Equation (\ref{fried1}), the signature   reversion phenomenon also exists and at this time it is only in the particular case of $5$-D brane-world scenario that the correction entropy term is positive.

\section{the generalized second law of thermodynamics}

 As was shown in various spacetimes in the literature, the identification\cite{Jacobson} of the gravitational field equations with the first law of thermodynamics on the apparent horizon is a general feature of the hypothetic holographic principle. The calculations presented above once again display the universal connection between them. It is of great interest to take a further step on the exploration of other thermodynamical aspects such as the tentative formulation of the thermodynamical second law in the settings of modified holographic dark energy model. Well after the elaboration on the corresponding unified first law of thermodynamics of the apparent horizon, it is not hard to compute the derivative of the the total entropy in the universe with respect to the cosmic time.

 The associated temperature on the apparent horizon can be expressed in the form
\begin{equation}
T_h=\frac{|\kappa|}{2\pi}=\frac{1}{2\pi \tilde{r}_A}(1-\frac{\dot{\tilde{r}}_A}{2H\tilde{r}_A})
\end{equation}
where $\dot{\tilde{r}}_A/(2H\tilde{r}_A)<1$ to ensure the positivity of the temperature.
Recognizing the entropy $S_h$ of the apparent horizon to be deduced through the connection between gravity and the first law of thermodynamics, we know that
\begin{equation}
T_h \dot{S}_h=\frac{1}{2\pi \tilde{r}_A}(1-\frac{\dot{\tilde{r}}_A}{2H\tilde{r}_A})[\frac{2\pi\tilde{r}_A \dot{\tilde{r}}_A}{G}+\frac{\pi d (N-5)(N-1)\Omega_{N-1}\tilde{r}^{N-2}_A \dot{\tilde{r}}_A}{6G_N}]; \label{ThSh}
\end{equation}
The two Friedmann equations~(\ref{fried1}) and ~(\ref{fried2}) can be recast into the following form
\begin{equation}
\rho_m+p_m=-2M_{pl}^2\dot{H}-\frac{d(N-5)(N-1)\Omega_{N-1}}{48\pi G_N}H^{3-N}\dot{H}.\label{Fried-ch1}
\end{equation}
As a result, Equation (~\ref{ThSh}) becomes
\begin{equation}
T_h \dot{S}_h=A(\rho_m+p_m)(1-\frac{\dot{\tilde{r}}_A}{2}).
\end{equation}
The entropy of matter fields inside the apparent horizon $S_m$ can be obtained by the Gibbs equation
\begin{equation}
T_m dS_m=d(\rho_m V)+p_m dV,
\end{equation}
where $E=\rho_m V$ is its energy and $p$ is its pressure in the horizon. Let us assume both perfect fluids are in equilibrium and
$T_h=T_m\equiv T$ throughout the evolution.Therefore, the variation of the total entropy can be recast into
\begin{equation}
T_h \dot{S}_h+T_m \dot{S}_m=T\dot{S}_h+V\dot{\rho}_m+(\rho_m+p_m)\dot{V}.\label{genetropy}
\end{equation}
With regard to Equation (~\ref{Fried-ch1}), we have
\begin{equation}
A(\rho_m+p_m)=8\pi M_{pl}^2 \dot{\tilde{r}}_A+\frac{d(N-5)(N-1)\Omega_{N-1}}{12 G_N}\tilde{r}_A^{N-3}\dot{\tilde{r}}_A
\end{equation}
and the last two terms of the right hand side of Equation(~\ref{genetropy}) read
\begin{equation}
T_m \dot{S}_m=4\pi (\rho_m+p_m)\tilde{r}_A^{2}(\dot{\tilde{r}}_A-1).\label{tmsm}
\end{equation}
Therefore, derived from the two Friedmann equations and Equations (\ref{continuation}) and (\ref{rA2}), the evolution of the total entropy can be obtained:
\begin{eqnarray}
\nonumber
T_h \dot{S}_h+T_m \dot{S}_m&=&\frac{1}{2}A(\rho_m+p_m)\dot{\tilde{r}}_A \\
&=&\frac{1}{2} A \tilde{r}_A^2(\rho_m+p_m)^2 [\frac{d(N-5)(N-1)\Omega_{N-1}}{48\pi G_N}H^{3-N}+\frac{1}{4\pi G}]^{-1}.\label{wholeentropy}
\end{eqnarray}
This fulfils the tentative formulation of the generalized second law of thermodynamics.

\section{Discussions and Conclusion}

The identification of the gravitational field equations with the first law of thermodynamics on the apparent horizon is a general feature of the hypothetic holographic principle. The calculations presented above once again display the universal connection between them. Well after the elaboration on the corresponding unified first law of thermodynamics of the apparent horizon, other thermodynamical aspects such as the tentative formulation of the thermodynamical second law in the settings of modified holographic dark energy model has also been explored.
From the entropy formula, there are several points to be emphasized on. The entropy formula consists of two parts: one corresponds to the contribution of 4-D scalar curvature, and the other counts on the extra-dimension effect in the bulk gravity, and it clearly reaffirms the existence of two gravity terms in our model. When $G\rightarrow \infty $, the first term of the corrected entropy formula vanishes and only the bulk gravity has effect on the area-entropy relation. Then the result of RS-II brane-world model can be recovered\cite{RS}. The situation is similar to the warped DGP model in brane-world scenario\cite{Shek1}\cite{Shek2}. Apart from the commonness to the general feature of DGP-like brane-world models, there are still some key difference in the formulation.

First, there is no factor $2$ in the numerator of the second term of the entropy formula, since there is no $Z_2$ symmetry to be respected, unlike some other DGP model constructions.

Second, in the entropy formula, the signature of the entropy correction term may be minus or plus, which depends on the number of the bulk extra dimensions. It's interesting to note that, the signature of the correction term will reverse well with the bulk dimension varying. Particularly, if there is an infinitely large extra spatial dimension with $N=n+3<5$, the bulk correction will be negative and tends to decrease the entropy of brane gravity. This feature is remarkably different from previous DGP brane-world model in the literature, which indicates that the dynamics of our present model is distinct from the latter in nature. Specifically speaking, although the first Friedmann equation appears to be of the same as that of the DGP model in $5-D$ spacetime, the thermodynamical analysis emphasizes on the fact that the situation would be otherwise. If the number of the extra dimensions is larger than two, the bulk correction will become positive again and it is interesting to find that the entropy correctness of bulk gravity will be influenced by the extra dimensions in this manner: decrease the brane entropy at low dimensions and increase it at high dimensions.

Third, as for the generalized second law of thermodynamics, the final expression in the last section above exhibits a simple new result which also makes some difference from the calculation on the original and warped DGP models. In present calculation, as the denominator in the formula of time-derivative of the total entropy is not positive-definite, the generalized second law of thermodynamics could be violated in certain cases. With the co-dimension two case $N=5$, the denominator is positive and just comes back to the LCDM model, and the generalized second law is trivially verified. When the number of the bulk spatial dimensions is larger than five, the generalized law can also be established in this way. The problematic situation lies in the previously most examined case with $N=4$ the $5-D$ embedding spacetime. The sign of the denominator is dependent on several constants and moreover the Hubble parameter. Notice that there exists the crossover scale $r_c$ in DGP brane-world. For five-dimensional space-time with $n=1$, the crossover scale is\cite{DGP}
$
r_c ~\sim~ {{M_{\rm Pl}^2}\over {M_*^3}}~.~$ Below the length scale $r_c$, we observe the four-dimensional gravity; while at distance $r > r_c$, we indeed observe high-dimensional gravity.
With this quantity included, the formula ~(\ref{wholeentropy}) in the case of $N=4$ can be rewritten as
$
 T_h \dot{S}_h+T_m \dot{S}_m=2\pi G A \tilde{r}_A^2(\rho_m+p_m)^2 (1-\frac{\Omega_3 d}{4 H r_c})^{-1}.
 $
If $H r_c\gg 1$, the above expression will be positive and the generalized law is satisfied.  For the other limit case $H r_c\ll 1$, the generalized law would not be verified.

 {\bf Acknowledgments.}
 The authors thank Prof. R. G. Cai and Prof. H. S. Zhang for valuable discussions and comments. H. Li is supported by National Foundation of China under grant No. 10747155 and 11205131. Y. Zhang is supported by the Ministry of Science and Technology of  the National Natural Science Foundation of China key project under grant Nos. 11175270, 11005164, 11073005 and 10935013, CQ CSTC under grant No.
 2010BB0408.

\end{document}